\documentclass[prl,aps,12pt]{revtex4}
\usepackage{dsfont}
\usepackage{amsmath}
\usepackage{amssymb}
\usepackage{epsfig}
\usepackage{amscd}
\usepackage{ulem}

\begin{document}

\title{Implementation of Liouville space search algorithm on strongly dipolar coupled nuclear spins}
\author{T. Gopinath and Anil Kumar}
\affiliation{{\small \it NMR Quantum Computing and Quantum Information Group.\\Department of Physics, and NMR Research Centre.\\
Indian Institute of Science, Bangalore - 560012, India.}}

\begin{abstract}
Liouville space search algorithm [Bruschweiler, Phys. Rev. Lett. {\bf 85}, 4815(2000).] utilizes mixed initial states of the ensemble, and has been successfully implemented earlier in weakly coupled spins, in which a spin can be 
identified as a qubit. It has recently been demonstrated that n-strongly coupled spins can be collectively treated as an n-qubit 
system. Application of algorithms in such systems, requires new approaches using transition selective pulses rather than 
qubit selective pulses. This work develops a modified version of Liouville space search algorithm, which is applicable for strongly 
as well as weakly coupled spins. 
All the steps of the algorithm, can be implemented by using transition selective pulses. 
Experimental implementation is carried out on a strongly dipolar coupled four qubit system.
\end{abstract}

\maketitle

\section{1. Introduction}

An important aspect of information processing is to search a specific data stored in a database \cite{5ic}. Classical computer 
operates on binary codes, any number or symbol is represented by combination of bits which can have the value either 
'0' or'1' \cite{5tur,5church}. 
To search an object from N objects, a classical computer must carry out O(N) operations \cite{5ic}. 
A quantum computer is based on the principle of quantum superposition, which offers the opportunity to perform parallel 
computations \cite{5deujoz,5div}. Grover search algorithm which uses quantum superposition, finds the marked object in O($\sqrt{N}$) 
operations \cite{5gr}. 
Most of the quantum algorithms, including Grover search algorithm, use a pure state (pseudo pure state in NMR) 
as an initial state \cite{5ic,5deujoz,5gr,5shor}. 

An alternative paradigm for computing has been suggested by Madi, Bruschweiler, and Ernst, which is based on the mixed states of 
ensemble, which are represented by density operators in the Liouville space \cite{5madi}.
Mixed states describe a statistical ensemble of spins, rather than individual spins, so that each element of the ensemble performs 
part of the computation, in the same way as each processor in a classical parallel computer \cite{5ic}. 
Bruschweiler proposed a novel search algorithm, that operates on mixed states in spin Liouville space,
and requires $log_2N$ queries to search a single object from N unsorted objects \cite{5bru}. 
The advantage of Liouville space computation is the use of an input state which is linear combination of different classical 
input states. Protopopescu had applied Liouville space search to the global optimization problem, and 
achieved exponential speed-up \cite{5dhe,5prot}. 
Liouville space search algorithm of Bruschweiler, considers only weakly coupled spins, in which mixed states are represented by linear combination 
of direct product states of spin polarization operators. The algorithm has been successfully implemented on weakly coupled systems by 
Xiao et al \cite{5xiao,5long,5xiao1,5long1}.

Most of the NMR Quantum information processing (QIP) experiments have utilized systems having indirect spin-spin couplings 
(scalar J couplings) \cite{5rev1,5rev2}. Since these couplings are mediated via the covalent bonds, the number of coupled spins and hence the number of 
qubits is limited to a few qubits. Another approach is to use direct dipole-dipole couplings between the spins which are larger in 
magnitude. In liquids the
dipolar couplings are averaged to zero due to rapid isotropic reorientations of the molecules and in rigid solids there are too many couplings yielding
broad lines \cite{5ern}. In molecules partially oriented in anisotropic media, like liquid crystals, one obtains partially averaged intra molecular dipolar
couplings with only a finite number of dipolar
coupled spins, yielding finite number of sharp NMR resonances \cite{5fung}. However in such cases often the dipolar couplings are large or comparable to chemical shifts
differences between the homonuclear coupled
spins, leading to spins becoming strongly coupled. The method of spin selective pulses and J evolution used in NMR-QIP of weakly coupled spins
(liquid state NMR-QIP), leads to complications in strongly coupled spins because of $\vec{I_i}.\vec{I_J}$ terms in the 
Hamiltonian \cite{5ern}. 
However it has been demonstrated that the $2^n$ non degenerate energy levels system can be collectively treated as an n-qubit system, 
similar to the case of quadrupolar nuclei oriented in liquid crystal \cite{5fung,5fung1,5mahesh}. The manipulation of the states of the qubit are then carried out 
by using transition selective pulses. Several gates and algorithms have been implemented in 
such systems \cite{5fung,5fung1,5mahesh,5mahesh1,5rana,5khitrin,5mahesh2}. 

In this work, we generalize the Liouville space search algorithm such that it can be implemented in strongly as well as weakly 
coupled spins. 
Experimental implementation is carried out on 
a four qubit system obtained by strongly dipolar coupled protons of 2-chloro, iodo benzene oriented in 
ZLI-1132 liquid crystal \cite{5mahesh1,5mahesh2}.
In section (2) we explain Liouville space search algorithm, in section (3) we explain the 
modified version, section (4) contains the experimental implementation, and conclusions are given in 
section (5).

\section{2. Liouville space search algorithm}

Liouville space search algorithm finds a marked object in $2^n$ unsorted objects, which correspond to $2^n$ eigen states \cite{5bru}. 
The eigen states in weakly coupled spins, are same as product states obtained from individual spin sates, and the corresponding 
density matrices are obtained from direct product 
of spin polarization operators \cite{5ern},

\begin{eqnarray}
\psi &=& \vert \alpha \beta \alpha ........\alpha \rangle 
= \vert \alpha \rangle \vert \beta \rangle \vert \alpha \rangle.......\vert \alpha \rangle  \nonumber \\
 \sigma &=& \vert \psi \rangle \langle \psi \vert 
=I_1^\alpha I_2^\beta I_3^\alpha .......I_n^\alpha, \label{5eqn1}
\end{eqnarray}

where $I_1^\alpha I_2^\beta I_3^\alpha .......I_n^\alpha$ is the direct product of polarization operators $I_k^{\alpha}$, $I_k^{\beta}$, 
defined as,

\begin{eqnarray}
I_k^\alpha = \vert \alpha_k \rangle \langle \alpha_k \vert = \frac{1}{2} (\mathds{1}_k +2I_{kz}) = 
\begin{pmatrix}  1&0 \cr 0&0  \end{pmatrix}  \nonumber \\ 
I_k^\beta = \vert \beta_k \rangle \langle \beta_k \vert = \frac{1}{2} (\mathds{1}_k -2I_{kz}) =
\begin{pmatrix}  0&0 \cr 0&1  \end{pmatrix}.  \label{5eqn2}
\end{eqnarray}

2$I_{kz}$ is the Pauli matrix $\sigma_z$ and $\mathds{1}_k$ is unity operator of the subspace of spin $I_k$. In Eq. (\ref{5eqn1}) 
$\sigma$ has the dimension 
of $2^n$ and it's matrix elements in the eigen basis (also known as Zeeman basis or product basis) are all zero except for one 
diagonal element belonging to the state 
$\vert \alpha \beta \alpha ........\alpha \rangle$ with coefficient 1, which represents the population of that state 
(Eq. \ref{5eqn1}) as one. $\vert \alpha \rangle$ and $\vert \beta \rangle$ are assigned to the 
computational 
basis states $\vert 0 \rangle$ and $\vert 1 \rangle$, thus Eq. (\ref{5eqn1}) can be written as,

\begin{eqnarray}
\psi &=& \vert 0 1 0 ........0 \rangle   \nonumber \\
 \sigma &=& \vert \psi \rangle \langle \psi \vert
=I_1^0 I_2^1 I_3^0 .......I_n^0.  \label{5eqn3}
\end{eqnarray}

The algorithm aims to search a single object in $N=2^n$ unsorted objects which are represented by $2^n$ basis states 
of 'n' qubits, which we call as work qubits. 
The density matrices of these states are represented in the Liouville space as,

\begin{eqnarray}
\sigma_1 &=&I_1^0 I_2^0 I_3^0 .......I_n^0 = \vert 0_10_20_3....0_n \rangle \langle 0_10_20_3....0_n \vert, \nonumber \\
\sigma_2 &=&I_1^0 I_2^0 I_3^0 .......I_{n-1}^0I_n^1=\vert 0_10_20_3....0_{n-1}1_n \rangle \langle 0_10_20_3....0_{n-1}1_n\vert 
\nonumber \\
&\vdots&  \nonumber \\
\sigma_{2^n} &=&I_1^1 I_2^1 I_3^1 .......I_n^1=\vert 1_11_21_3....1_n \rangle \langle 1_11_21_3....1_n \vert. \label{5eqn4}
\end{eqnarray}

Let the marked state is $\sigma_m$ (m=1,2,....,$2^n$ ). 
The algorithm uses an extra qubit known as ancilla qubit $I_0$, and finds the marked object in 'n' ($=log_2N$) queries.

{\bf (i) Preparation of initial states:} 'n' initial states, $\rho_{in}^k$ (k=1, 2,...,n), are prepared in 'n' different experiments, 
$\rho_{in}^k$ is given by \cite{5bru,5xiao},

\begin{eqnarray}
\rho^k_{in} &=&  I_0^{0} I_k^0 \nonumber \\
&=&  I_0^{0} [\mathds{1}_1\mathds{1}_2.....\mathds{1}_{k-1}I_k^{0}\mathds{1}_{k+1}......\mathds{1}_n] \nonumber \\
&=&I_0^{0} [(I_1^0I_2^0....I_{k-1}^0I_k^0I_{k+1}^0.....I_n^0)+(I_1^0I_2^0....I_{k-1}^0I_k^0I_{k+1}^0.....I_n^1)+........... \nonumber \\
&& +(I_1^1I_2^1....I_{k-1}^1I_k^0I_{k+1}^1.....I_n^1)] \nonumber \\
&=&I_0^{0}[\sigma_1^k+\sigma_2^k+..........+\sigma_{2^{n-1}}^k]=I_0^{0}\sigma_{in}^k.  \label{5eqn5}
\end{eqnarray}

In Eq. (\ref{5eqn5}), ancilla qubit and
$k^{th}$ work qubit are in state $\vert 0 \rangle$, whereas each of the remaining $(n-1)$ qubits are in state $\vert 0 \rangle$ or
$\vert 1 \rangle$. Thus $\rho_{in}^k$ has equal and non-zero populations (each of magnitude 1) in $2^{n-1}$ energy 
levels \cite{5bru,5xiao}. In other words, all the matrix elements of $\rho_{in}^k$ are zero, except for $2^{n-1}$ diagonal 
elements, each of which represent populations of value 1.
The initial state $\rho_{in}^k$ can be prepared in weakly coupled spins by using spin selective pulses and J-evolutions \cite{5xiao,5long}.

In Eq. (\ref{5eqn5}) since $k^{th}$ work qubit of each $\sigma_i^k$, is already set to $\vert 0_k \rangle$, if 
$\sigma_{in}^k$ contains the marked state 
$\sigma_m$, then it implies that $k^{th}$ work qubit of the marked state is 
$\vert 0_k \rangle$. In other words $\sigma_m$= $I_1^{0/1}I_2^{0/1}....I_{k-1}^{0/1} {\bf I_k^{0}}I_{k+1}^{0/1}....I_n^{0/1}$. 
If $\sigma_{in}^k$ does not contain $\sigma_m$, 
then $k^{th}$ work qubit of the marked state must be 
$\vert 1_k \rangle$, that is $\sigma_m$= $I_1^{0/1}I_2^{0/1}....I_{k-1}^{0/1}{\bf I_k^{1}}I_{k+1}^{0/1}....I_n^{0/1}$ \cite{5xiao,5long}.

{\bf (ii) Oracle:} 
Oracle is a black box, contents of which are unknown to the reader. 
Unitary operator (U) of oracle, inverts the states $I_0^0\sigma_m$ and $I_0^1\sigma_m$. 
The resultant state $\rho_f^k$ is then given by,

\begin{eqnarray}
\rho^k_{f}=U(\rho^k_{in}) = U(I_0^{0} \sigma_{in}^k) 
&=& I_0^0 \sigma_{in}^k, \mbox{if $\sigma_{in}^k$ does not contain the marked state $\sigma_m$, $I_k$ $\Rightarrow$ $I_k^1$}   \nonumber \\
&=& I_0^1\sigma_m + I_0^0[\sigma_{in}^k-\sigma_m], \nonumber \\
&&\mbox{if $\sigma_{in}^k$ contains the marked state $\sigma_m$, $I_k$ $\Rightarrow$ $I_k^0$}.  \label{5eqn7}
\end{eqnarray}

As one can see in Eq. (\ref{5eqn7}), 'U' is operated simultaneously on all the qubits of the ensemble. 

{\bf (iii) Measurement:}
The ancilla qubit ($I_0$) is detected by applying a $(\pi/2)_y^0$ pulse on $\rho_f^k$ \cite{5xiao}. 
The spectrum of $I_0$ contains $2^{n-1}$ peaks corresponding to $2^{n-1}$ states of $\sigma_{in}^k$. 
In Eq. (\ref{5eqn7}), if $\rho_f^k$ = $\rho_{in}^k$ then all the peaks are of positive intensity, which indicates that the $k^{th}$ qubit of the marked state 
is in state $\vert 1 \rangle$ or $\sigma_m$= $I_1^{0/1}I_2^{0/1}....I_{k-1}^{0/1}{\bf I_k^{1}}I_{k+1}^{0/1}....I_n^{0/1}$ 
\cite{5bru,5xiao}. 
If $\rho_f^k$ = $I_0^1\sigma_m-I_0^0[\sigma_{in}^k-\sigma]$, then the spectrum contains a negative peak corresponding to $\sigma_m$, 
whereas remaining $(2^{n-1}-1)$ peaks are of positive intensity, which indicates that 
$\sigma_m$= $I_1^{0/1}I_2^{0/1}....I_{k-1}^{0/1}{\bf I_k^{0}}I_{k+1}^{0/1}....I_n^{0/1}$ \cite{5xiao}.

A sequence of 'n' experiments with input $\rho_{in}^k=I_0^0I_k^0$, k=1, 2, ...., n, are recorded. 
In the $k^{th}$ experiment if all the peaks 
are positive then the marked state contains $I_k^1$, otherwise it is $I_k^0$ \cite{5bru}. 
Thus in 'n' experiments one can find the states of all the work qubits, $I_1$, $I_2$,....., $I_n$ of the marked state.

\section{3. Generalized Liouville space search algorithm}

Liouville space search algorithm has been implemented in NMR by using weakly J-coupled spins \cite{5xiao,5long}. 
Various steps of the algorithm are implemented by using qubit (spin) selective pulses and evolving the system under 
J-couplings \cite{5xiao,5long}. 
It is difficult to implement this algorithm in strongly coupled spins, since the spins are not 
individually addressable in the spectrum. Also it is difficult to manipulate the time evolution under a desired Hamiltonian. In other 
words eigen states are not 
the product states of individual spins. 
However an n-coupled system can be collectively treated as an n-qubit system, by treating the $2^n$ eigen states as computational 
basis states \cite{5mahesh,5mahesh1,5rana,5khitrin}. 
In such systems it is difficult to apply qubit selective pulses, but one can still manipulate the states by applying
transition selective pulses. Thus the original algorithm explained in section (2), is modified in such a way that various
steps of the algorithm can be implemented by using transition selective pulses.
The modified algorithm is implemented here in strongly coupled spins, but can be implemented in weakly coupled spins as well.

Consider an  (n+1)-qubit system (weakly or strongly coupled), where the $2^{n+1}$ eigen states are labeled as 
$\vert 0_00_1....0_n \rangle$ .......$\vert 1_01_1....1_n \rangle$, where the subscripts '0' and 1, 2...., n represent ancilla qubit and 
'n' work qubits respectively. 
The search space contains $2^n$ states of 'n' work qubits, whose density matrices in $2^n$- dimensional subspace of $2^{n+1}$- dimensional 
eigen basis, are given by,

\begin{eqnarray}
\sigma_{1}&=& \vert \psi_1 \rangle \langle \psi_1 \vert = \vert 0_10_2.....0_n \rangle \langle 0_10_2.....0_n \vert \nonumber \\
\sigma_{2}&=& \vert \psi_2 \rangle \langle \psi_2 \vert=\vert 0_10_2.....1_n \rangle \langle 0_10_2.....1_n \vert  \nonumber \\
&\vdots&   \nonumber \\
\sigma_{2^n}&=& \vert \psi_{2^n} \rangle \langle \psi_{2^n} \vert=\vert 1_11_2.....1_n \rangle \langle 1_11_2.....1_n \vert. \label{5eqn12}
\end{eqnarray}

In Eq. (\ref{5eqn12}), the matrix elements of $\sigma_i$ ($i = 1, 2,....,2^n$) are zero, except for one diagonal element of value 1, 
which corresponds to the state $\psi_i$. 
The marked state is represented by,
\begin{eqnarray}
\sigma_{m}=\vert \psi_m \rangle \langle \psi_m \vert=\vert x_1x_2......x_n \rangle \langle x_1x_2.....x_n \vert, 
\mbox{where m=1, 2,..., $2^n$ and $x_i$=0 or 1}. \label{5eqn13}
\end{eqnarray}

{\bf(i) Preparation of initial states:} 'n' initial states $\rho_{in}^k$ (k=1, 2, ...,n) are prepared in 'n' different 
experiments using the method of POPS (pair of pseudopure states) \cite{5fung}. 
POPS(i, j) contains populations only in states $\vert i \rangle$ and $\vert j \rangle$, given by 'p' and '-p' respectively, 
where 
the value of 'p' depends on the details of preparation \cite{5fung}. POPS(i, j) can be achieved by inverting the populations of
a pair of states $\vert i \rangle$ and $\vert j \rangle$ and subtracting the resultant population distribution from that of 
equilibrium state \cite{5fung}.

In each initial state $\rho_{in}^k$, ancilla and $k^{th}$ work qubits are in state $\vert 0 \rangle$, whereas each of the remaining (n-1) 
qubits are in state 0 or 1. Thus the initial state contains non zero populations in $2^{n-1}$ states, also for every state of population 
$p_i^k$ there exist another state of population of $-p_i^k$. 
In other words, $\rho_{in}^k$ is a sum of $(2^{n-1}/2)$ POPS, thus it represents a highly mixed state. $\rho_{in}^k$ is prepared by subtracting $\rho_k$ from 
$\rho_{eq}$ (equilibrium populations), where $\rho_k$ is obtained by inverting the populations of
$(2^{n-1}/2)$ pairs of states, and subtracting the resultant population distribution from that of equilibrium state, $\rho_{in}^k$ is
given by,

\begin{eqnarray}
\rho_{in}^k&=&\rho_{eq}-\rho_k, \nonumber \\
&=&  p_1^k\vert {\bf 0_0}0_10_2....0_{k-1}{\bf 0_{k}}0_{k+1}....0_n \rangle \langle {\bf 0_0}0_10_2.....{\bf 0_k}.....0_n \vert \nonumber \\
&& +p_2^k\vert {\bf 0_0}0_10_2....0_{k-1}{\bf 0_{k}}0_{k+1}....1_n \rangle \langle {\bf 0_0}0_10_2.....0_{k-1}{\bf 0_k}0_{k+1}.....1_n \vert    \nonumber \\
&&+.............+p_{2^{n-1}}^k\vert {\bf 0_0}1_11_2....1_{k-1}{\bf 0_{k}}1_{k+1}....1_n \rangle \langle {\bf 0_0}1_11_2.....1_{k-1}
{\bf 0_k}1_{k+1}.....1_n \vert
\nonumber \\
&=& p_1^k \vert 0_0\psi_1^k \rangle \langle 0_0 \psi_1^k \vert + p_2^k \vert 0_0\psi_2^k \rangle \langle 0_0 \psi_2^k \vert +........+
p_{2^{n-1}}^k \vert 0_0\psi_{2^{n-1}}^k \rangle \langle 0_0 \psi_{2^{n-1}}^k \vert, \label{5eqn14}
\end{eqnarray}

where $p_1^k, p_2^k....p_{2^{n-1}}^k$ represent the populations in the respective levels, and for every $p_i^k$ there 
exists a -$p_i^k$, thus $\rho_{in}^k$ represents a sum of $(2^{n-1}/2)$ POPS.
In Eq. (\ref{5eqn14}) the $k^{th}$ work qubit is set in state $\vert 0_k \rangle$, hence if $\rho_{in}^k$ contains the marked state 
$\vert \psi_m \rangle=\vert x_1x_2....x_k....x_{n} \rangle$, it implies that 
the $k^{th}$ work qubit of $\vert \psi_m \rangle$ that is $x_k=0$, whereas if $\rho_{in}^k$ does not contain 
$\vert \psi_m \rangle$ then $x_k=1$.

{\bf(ii) Oracle:} 
The unitary operator of the oracle acting on $\rho_{in}^k$, flips the states $\vert 0_0\psi_m \rangle$ and $\vert 1_0\psi_m \rangle$. 
'U' can be implemented by applying a selective $(\pi)$ pulse on transition 
$\vert 0_0\psi_m \rangle$ $\Leftrightarrow$  $\vert 1_0\psi_m \rangle$. The final state $\rho_f^k$ is 
given by,

\begin{eqnarray}
\rho^k_{f}=U(\rho^k_{in}) &=& \rho^k_{in}, \mbox{if $\rho_{in}^k$ does not contain the marked state 
$\vert \psi_m \rangle$ $\Rightarrow$ $x_k$=1}   \nonumber \\ 
&=&p_m^k \vert 1_0 \psi_m \rangle \langle 1_0 \psi_m \vert + (\rho_{in}^k-p_m^k \vert 0_0 \psi_m \rangle \langle 0_0 \psi_m \vert), \nonumber \\
&& \mbox{if $\rho_{in}^k$ contains the marked state $\vert \psi_m \rangle$ $\Rightarrow$ $x_k$=0}. \label{5eqn15}
\end{eqnarray}

{\bf (iii) Measurement:}
A multi frequency $(\pi/2)_y$ pulse is applied on $2^{n}$ unconnected transitions (ancilla qubit transitions) corresponding to 
$2^n$ pairs of states 
$\vert 0_0i_1i_2........i_n \rangle \Leftrightarrow  \vert 1_0i_1i_2........i_n \rangle$, 
where $i_1, i_2,....i_n$ = 0 or 1 , followed by the detection. 
In case of weakly coupled system with conventional labeling (that is the states 
$\vert \alpha \rangle$ and $\vert \beta \rangle$ of each spin, 
represent $\vert 0 \rangle$ and $\vert 1 \rangle$ respectively), these 
$2^n$ transitions belong to $I_0$ qubit. Hence the read-out pulse is selective $(\pi/2)_y$ pulse on $I_0$ qubit. 
Thus in case of weakly coupled spins, measurement strategy is same as that of the original 
algorithm \cite{5bru,5xiao}, [section 2].

In Eq. (\ref{5eqn15}), if the final state $\rho_{f}^k$=$\rho_{in}^k$, then the MF pulse converts the populations 
$p_1^k \vert 0_0 \psi_1^k \rangle \langle 0_0 \psi_1^k \vert$, 
$p_2^k \vert 0_0 \psi_2^k \rangle \langle 0_0 \psi_2^k \vert$,......
$p_{2^{n-1}}^k \vert 0_0 \psi_{2^{n-1}}^k \rangle \langle 0_0 \psi_{2^{n-1}}^k \vert$, to 
$2^{n-1}$ coherences. Since for every $p_i^k$ there exists -$p_i^k$, number of positive transitions (associated with positive $p_i^k$) is 
equal to number of negative transitions (associated with negative $p_i^k$). 
If $\rho_f^k$ contains the marked state (corresponds to second equality of Eq. \ref{5eqn15}), then positive and negative populations 
(other than $p_m^k\vert 1_0 \psi_m \rangle \langle 1_0 \psi_m \vert$) are converted in to positive and 
negative transitions respectively, whereas a positive or negative $p_m^k \vert 1_0 \psi_m \rangle \langle 1_0 \psi_m \vert$ gives 
a negative or positive peak respectively. 
Thus, if $\rho_f^k$ contains the marked state, then spectrum contains $(2^{n-1}/2)\pm 1$ positive peaks and $(2^{n-1}/2)\mp 1$ negative 
peaks, where + and - respectively corresponds to negative and positive $p_m^k$.
 
In summary, in the $k^{th}$ experiment with initial state $\rho_{in}^k$, 
if the spectrum of final state contains equal number of positive and negative peaks, 
then the $k^{th}$ work qubit is in state $\vert 1 \rangle$ or $x_k=1$. 
Whereas if the spectrum contains unequal number of positive and negative peaks then $x_k=0$.
This procedure is repeated for each of the 'n' initial states $\rho_{in}^k$ (k=1,2...n), then one obtains the marked state 
$\sigma_m= \vert x_1x_2......x_n \rangle \langle x_1x_2......x_n \vert$.

It is to be noted that, since $\rho_{in}^k=\rho_{eq}-\rho_{k}$ (Eq. \ref{5eqn14}), implementation of 'U' and measurement are 
performed in different experiments for $\rho_{eq}$ and $\rho_{k}$ (k = 1, 2,....,n). Now the spectrum of $\rho_f^k$ is obtained 
by subtracting the final spectrum of $\rho_{k}$ from that of $\rho_{eq}$. Thus the total number of experiments is (n+1).

\section{4. Experimental implementation on a strongly dipolar coupled 4-qubit system}

The system chosen is, four protons of 2-chloro, iodo benzene oriented in ZLI-1132 liquid crystal \cite{5mahesh1}. 
The four protons are strongly coupled to each other through dipolar couplings, and contains 16 eigen states. The total Hamiltonian of 
the system contains, 
chemical shift ($H_c$), dipolar ($H_{DD}$), and J-coupling ($H_{J}$) terms, given by,

\begin{eqnarray}
H=H_c+H_{DD}+H_J=\sum_{i=1}^4 \omega_i I_{iz}+ \sum_{i,j=1,and i<j}^4 [2 \pi D_{ij} (3I_{iz}I_{jz}-I_i.I_j)+2 \pi J_{ij} I_i.I_j]. \label{5eqn16}
\end{eqnarray}

\begin{figure}
\begin{center}
\epsfig{file=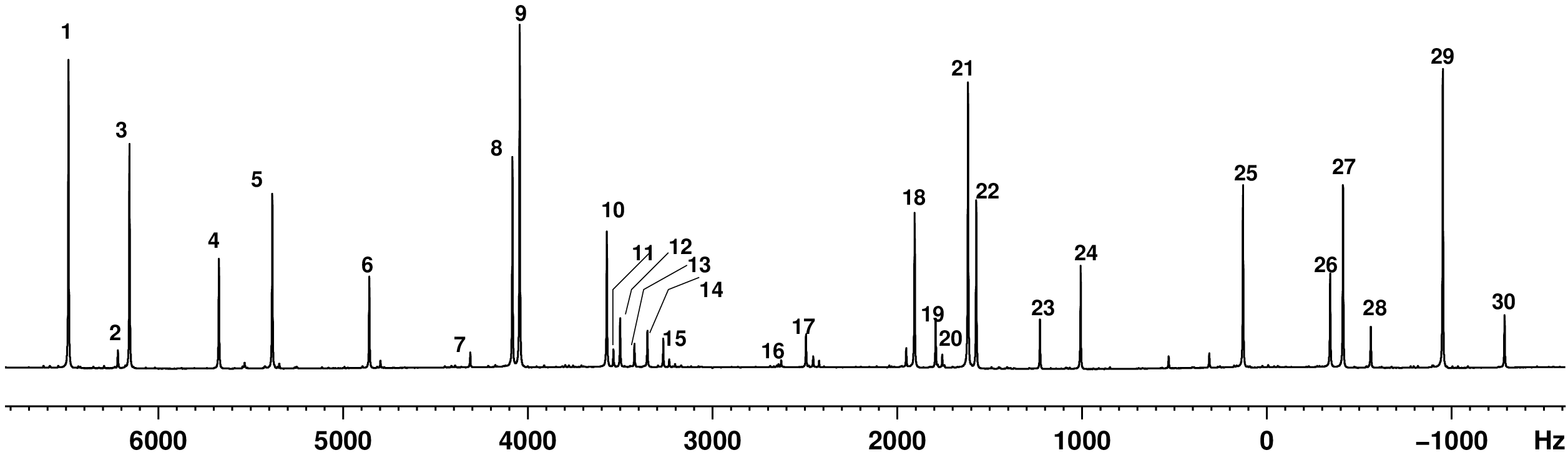,width=5cm,angle=270}
\caption{$^{1}H$ equilibrium spectrum of 2-chloro, iodo benzene oriented in ZLI-1132 liquid crystal, recorded at room temperature (300k) 
on a 500 MHz NMR spectrometer. The four protons of the molecule are strongly dipolar coupled to each other.
Various transitions are numbered in increasing order from left to right (decreasing frequency).}
\label{5equi}
\end{center}
\end{figure}

\begin{figure}
\begin{center}
\epsfig{file=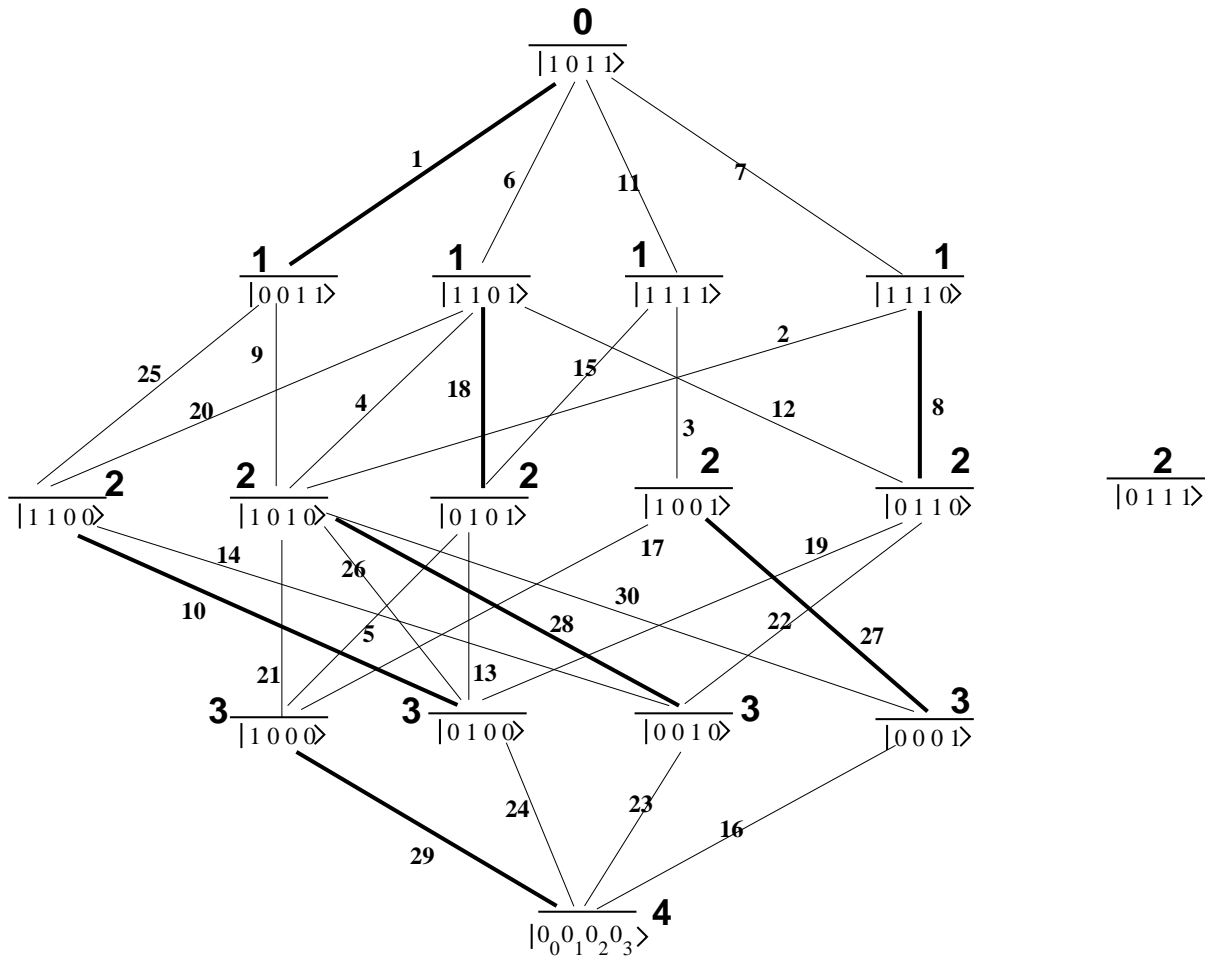,width=16cm,angle=0}
\caption{Energy level diagram of oriented 2-chloro, iodo benzene, obtained by investigating the connectivity of
various transitions of Fig. \ref{5equi} by using a Z-COSY experiment \cite{5mahesh1}. The equilibrium populations ($\rho_{eq}$) of
various levels
(represented by bold numbers)
are proportional to the Zeeman energies of the protons, schematically represented by the energy level diagram. 
The 16 eigen states are labeled as basis states of a four qubit system. The strategy adopted for this labeling scheme is explained 
in the text. The transitions represented by dark lines,
correspond to ancilla (zeroth) qubit.}
\label{energy}
\end{center}
\end{figure}

Experiments are carried out on AV-500 NMR spectrometer. 
Figure (\ref{5equi}) shows the $^{1}H$ 1D spectrum obtained by applying a hard $(\pi/2)_y$ pulse. 
In order to perform the Liouville search, one needs to know the pair of states from which each transition originates or in other words, 
we need to know the complete energy level diagram of the spin system. The energy level diagram of this system, has been obtained 
by Mahesh et al, using the connectivity information from Z-COSY experiment \cite{5mahesh1}. 
The pulse sequence of a Z-COSY experiment is, 
90- $t_1$ - $\alpha$ - $\tau$ - $\beta$ - $t_2$, where only the longitudinal
magnetization is retained during the interval $\tau$ by either
phase cycling or by a gradient pulse \cite{5grace,5rana}. The $90^o$ pulse
converts the equilibrium z-magnetization into coherences,
which are frequency labeled during the period $t_1$.
The small angle $\alpha$ pulse ensures that each cross-section
parallel to $\omega_2$ from the resulting 2D spectrum is equivalent
to a one-dimensional (1D) experiment in which the
peak corresponding to the diagonal is selectively inverted \cite{5grace}.
The directly connected transitions to the inverted
transition are finally measured in the linear regime by a
small angle $\beta$ pulse. 
Figure (\ref{energy}) shows the energy level diagram of this system, obtained from Z-COSY experiment \cite{5mahesh1}, where each of the 
transitions of Fig. (\ref{5equi}) are assigned to a pair of energy levels.

The 16 eigen states (Fig. \ref{energy}) 
are labeled as 
computational 
basis states of a four qubit system, given by, $\vert 0_00_10_20_3 \rangle$, $\vert 0_00_10_21_3 \rangle$, .........
$\vert 1_01_11_21_3 \rangle$, where the subscripts '0' and 1, 2, 3 represent ancilla qubit and three work qubits. 
It is to be noted that the labeling of eigen states shown in Fig. (\ref{energy}), is different from that of ref. \cite{5mahesh1}. 
The labeling in our case follows two strategies: 
(i) Find eight unconnected transitions, and label the eigen states such that these transitions belong to ancilla qubit. In other words, each of the eight 
transitions is associated with a pair of states $\vert 0_0i_1i_2i_3 \rangle$ $\Leftrightarrow$ $\vert 1_0i_1i_2i_3 \rangle$, 
where $i_1i_2i_3 = 000, 001,....., 111$; the states  $\vert 0_0i_1i_2i_3 \rangle$ and $\vert 1_0i_1i_2i_3 \rangle$ respectively represent the lower and 
upper states of the transition.  
However in this system we could find only seven unconnected transitions, represented by dark lines in Fig. (\ref{energy}),
the transition between the states $\vert 0_01_11_21_3 \rangle$ $\Leftrightarrow$ $\vert 1_01_11_21_3 \rangle$ is not observed due to low intensity.
(ii) The initial states $\rho_{in}^k$ (Eq. \ref{5eqn18}) contain the eigen states
$\vert 0000 \rangle$, $\vert 0001 \rangle$, $\vert 0010 \rangle$,
$\vert 0100 \rangle$, $\vert 0011 \rangle$ and $\vert 0101 \rangle$, hence to prepare $\rho_{in}^k$ each of these states should not be 
isolated 
from the rest of eigen states, for example the labeling of any of these eigen states can not be interchanged with that of 
$\vert 0111 \rangle$ since this state is isolated (Fig. \ref{energy}). 
Following these two criteria there exists many possibilities of labeling these eigen states, the present being 
one of them. It may be pointed out that the second strategy is not necessary, if none of the states are isolated.

Liouville space search is now performed, following the procedure given in section (3). 
The search space contains eight states of the work qubits, given by,

\begin{eqnarray}
\sigma_{1}&=&\vert \psi_1 \rangle \langle \psi_1 \vert = \vert 0_10_20_3 \rangle \langle 0_10_20_3 \vert, \hspace{0.3cm} 
\sigma_{2}=\vert \psi_2 \rangle \langle \psi_2 \vert=\vert 0_10_21_3 \rangle \langle 0_10_21_3 \vert  \nonumber \\
\sigma_{3}&=&\vert \psi_3 \rangle \langle \psi_3 \vert = \vert 0_11_20_3 \rangle \langle 0_11_20_3 \vert, \hspace{0.3cm}
\sigma_{4}=\vert \psi_4 \rangle \langle \psi_4 \vert=\vert 0_11_21_3 \rangle \langle 0_11_21_3 \vert  \nonumber \\
\sigma_{5}&=&\vert \psi_5 \rangle \langle \psi_5 \vert = \vert 1_10_20_3 \rangle \langle 1_10_20_3 \vert, \hspace{0.3cm}
\sigma_{6}=\vert \psi_6 \rangle \langle \psi_6 \vert=\vert 1_10_21_3 \rangle \langle 1_10_21_3 \vert  \nonumber \\
\sigma_{7}&=&\vert \psi_7 \rangle \langle \psi_7 \vert = \vert 1_11_20_3 \rangle \langle 1_11_20_3 \vert, \hspace{0.2cm} and \hspace{0.2cm}
\sigma_{8}=\vert \psi_8 \rangle \langle \psi_8 \vert=\vert 1_11_21_3 \rangle \langle 1_11_21_3 \vert.  \nonumber \\  \label{5eqn17}
\end{eqnarray}

Let the marked state is $\sigma_6=\vert \psi_6 \rangle \langle \psi_6 \vert$. 

{\bf (i) Preparation of initial states:}

Three initial states $\rho_{in}^k$ (k=1, 2, 3) are required since there are three work qubits. 
From Eq. (\ref{5eqn14}) we can write these initial states as,

\begin{eqnarray}
\rho_{in}^1&=&p_1^1 \vert 0_00_10_20_3 \rangle \langle 0_00_10_20_3 \vert+p_2^1\vert 0_00_10_21_3 \rangle \langle 0_00_10_21_3 \vert
\nonumber \\
&&+p_3^1\vert 0_00_11_20_3 \rangle \langle 0_00_11_20_3 \vert+p_4^1\vert 0_00_11_21_3 \rangle \langle 0_00_11_21_3 \vert  \nonumber \\
\rho_{in}^2&=&p_1^2\vert 0_00_10_20_3 \rangle \langle 0_00_10_20_3 \vert+ p_2^2\vert 0_00_10_21_3 \rangle \langle 0_00_10_21_3 \vert
\nonumber \\
&&+p_3^2\vert 0_01_10_20_3 \rangle \langle 0_01_10_20_3 \vert+p_4^2\vert 0_01_10_21_3 \rangle \langle 0_01_10_21_3 \vert  \nonumber \\
\rho_{in}^3&=&p_1^3 \vert 0_00_10_20_3 \rangle \langle 0_00_10_20_3 \vert+ p_2^3\vert 0_00_10_21_3 \rangle \langle 0_00_10_21_3 \vert
\nonumber \\
&&+p_3^3\vert 0_01_10_20_3 \rangle \langle 0_01_10_20_3 \vert+p_4^3\vert 0_01_10_21_3 \rangle \langle 0_01_10_21_3 \vert. \label{5eqn18}
\end{eqnarray}

In each experiment, for every $p_i^k$ there exists another population which is -$p_i^k$.

\begin{figure}
\begin{center}
\epsfig{file=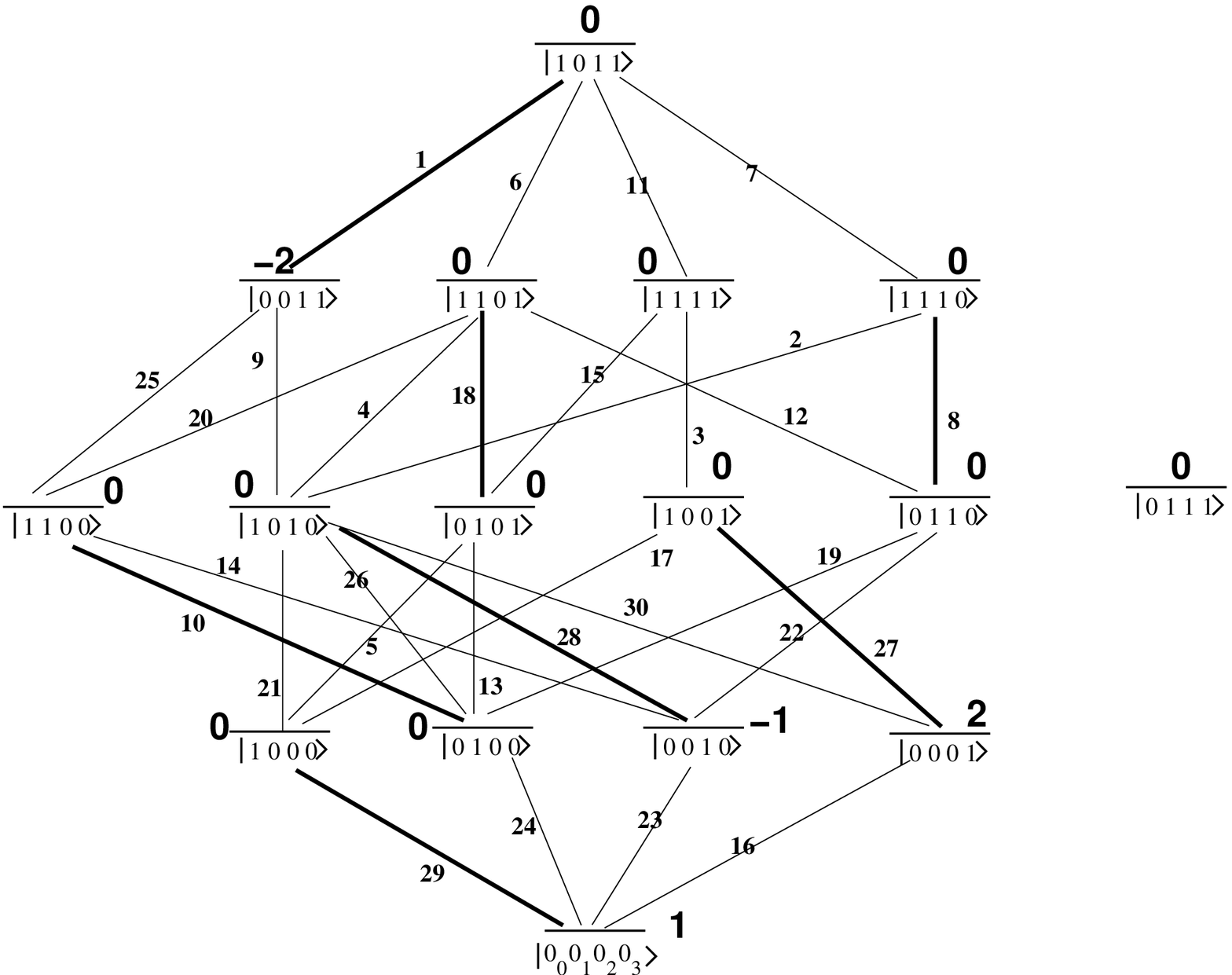,width=16cm,angle=0}
\caption{Energy level diagram of Fig. (\ref{energy}), with the populations of various states
corresponding to $\rho_{in}^1$ (Eq. \ref{5eqn19}). The transitions represented by dark lines, correspond to ancilla qubit.}
\label{in1}
\end{center}
\end{figure}

A density matrix $\rho_1$ is prepared from $\rho_{eq}$ (Fig. \ref{energy}), 
by applying a selective $\pi$ pulse on transition (23) which interchanges the populations of $\vert 0_00_10_20_3 \rangle$ $\&$
$\vert 0_00_11_20_3 \rangle$, and a cascade of $\pi$ pulses on (30), (9) and (30) which interchanges the populations of 
$\vert 0_00_10_21_3 \rangle$ $\&$ $\vert 0_00_11_21_3 \rangle$.  
Now $\rho_{in}^1$ is obtained by subtracting $\rho_1$ from $\rho_{eq}$, the resultant populations of 
$\rho_{in}^1$ are shown in Fig. (\ref{in1}), which can also be written as,

\begin{eqnarray}
\rho_{in}^1=\rho_{eq}-\rho_1&=&\vert 0_00_10_20_3 \rangle \langle 0_00_10_20_3 \vert+2\vert 0_00_10_21_3 \rangle \langle 0_00_10_21_3 \vert \nonumber \\
&&-\vert 0_00_11_20_3 \rangle \langle 0_00_11_20_3 \vert-2\vert 0_00_11_21_3 \rangle \langle 0_00_11_21_3 \vert. \label{5eqn19}
\end{eqnarray}

\begin{figure}
\begin{center}
\epsfig{file=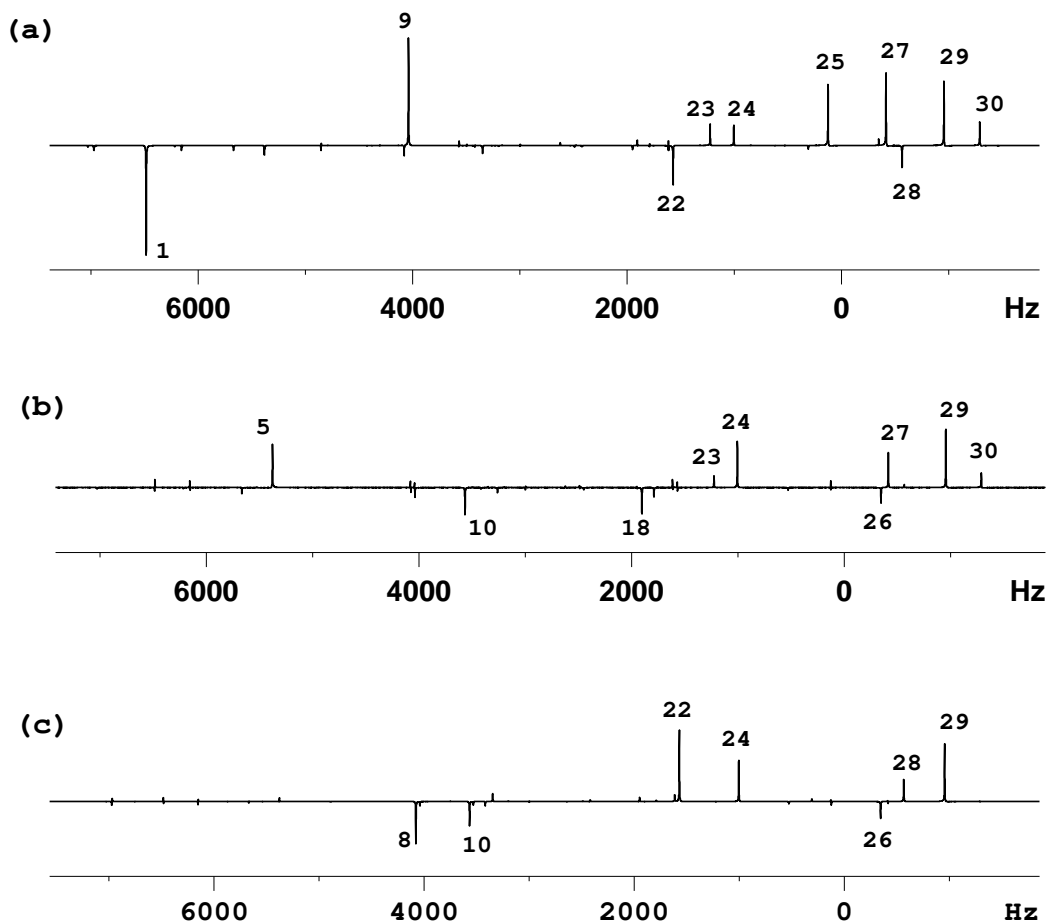,width=14cm,angle=0}
\caption{The spectra of (a), (b) and (c) respectively represent the population distributions of $\rho_{in}^1$ (Fig. \ref{in1}),
$\rho_{in}^2$ (Fig. \ref{in2}) and $\rho_{in}^3$ (Fig. \ref{in3}). (a), (b) and (c) are respectively obtained by subtracting the spectra
(recorded with $5^o$ pulse ) of $\rho_{1}$, $\rho_{2}$ and $\rho_{3}$ from equilibrium ($\rho_{eq}$) spectrum
(also recorded with a $5^o$ pulse). Preparation of $\rho_1$, $\rho_2$ and $\rho_3$ are explained in the text. Some small intensities 
such as 14 $\&$ 16 in (a), are not observed/marked.}
\label{ppops}
\end{center}
\end{figure}

Figure (\ref{ppops}a) shows the spectrum of $\rho_{in}^1$ (Fig. \ref{in1}), obtained by subtracting the spectrum of $\rho_1$ 
from $\rho_{eq}$, 
where each of the spectra of $\rho_1$ and $\rho_{eq}$ were recoded by applying a $5^o$ pulse. 
In Fig. (\ref{in1}), the population difference corresponding to transitions 
(29), (24), (23), (27), (30), (25) and (9) are positive, while those of (16), (22), (28), (14) and (1) are negative. 
This is confirmed from the spectrum of Fig. (\ref{ppops}a), the transitions (14) and (16) are not visible due to low intensity.

\begin{figure}
\begin{center}
\epsfig{file=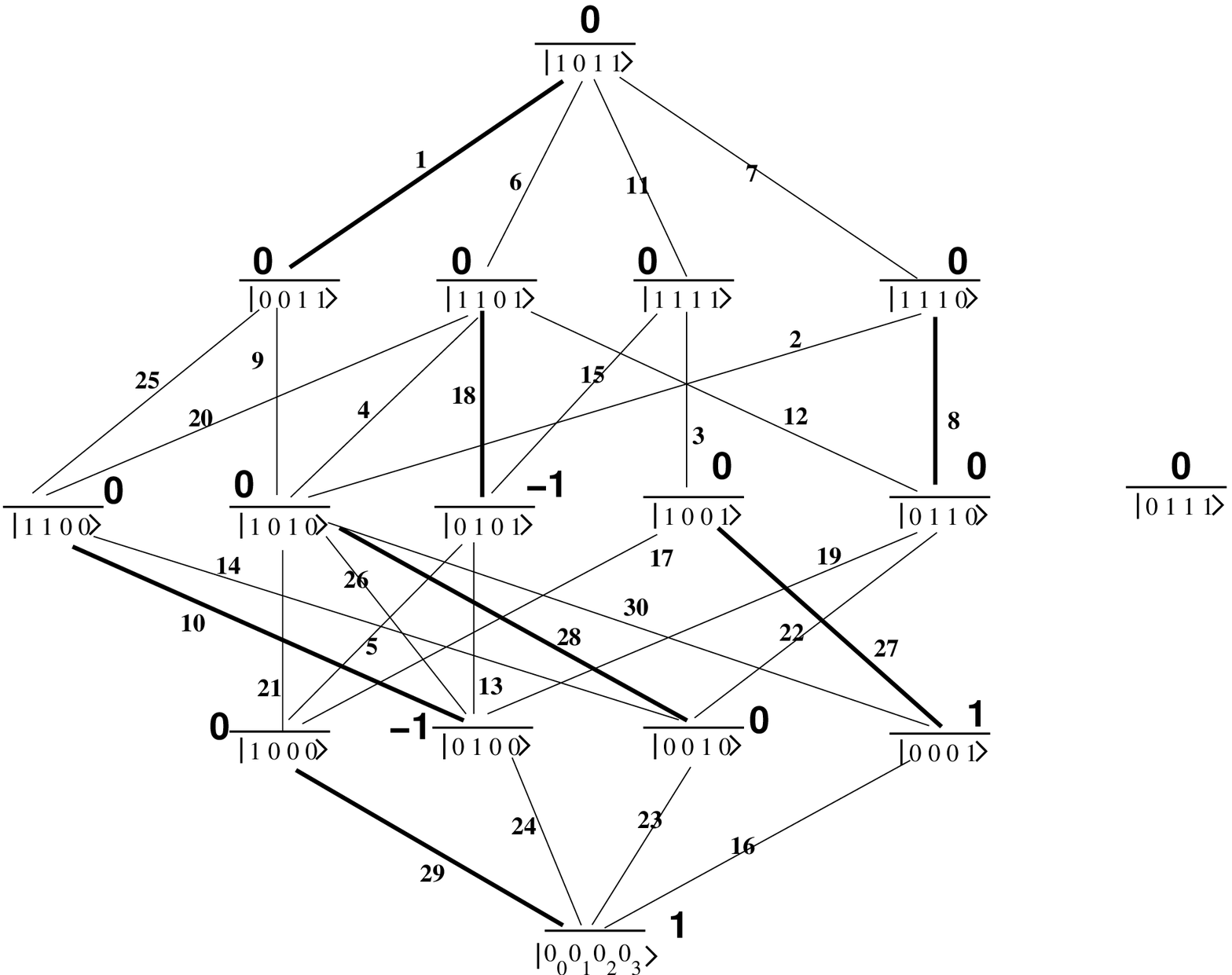,width=16cm,angle=0}
\caption{Energy level diagram of Fig. (\ref{energy}), with the populations of various states
corresponding to $\rho_{in}^2$ (Eq. \ref{5eqn20}), and the transitions represented by dark lines, correspond to ancilla qubit}
\label{in2}
\end{center}
\end{figure}

The $\rho_{in}^2$ is given by, 

\begin{eqnarray}
\rho_{in}^2=\rho_{eq}-\rho_2&=&\vert 0_00_10_20_3 \rangle \langle 0_00_10_20_3 \vert+
\vert 0_00_10_21_3 \rangle \langle 0_00_10_21_3 \vert   \nonumber \\
&& -\vert 0_01_10_20_3 \rangle \langle 0_01_10_20_3 \vert-\vert 0_01_10_21_3 \rangle \langle 0_01_10_21_3 \vert, \label{5eqn20}
\end{eqnarray}

where $\rho_2$ is obtained from $\rho_{eq}$ (Fig. \ref{energy}), 
by applying six selective $\pi$ pulses on transitions (24), (30), (21), (5), (21) and (30) respectively, where the first pulse inverts the 
populations of $\vert 0_00_10_20_3 \rangle$ $\&$ $\vert 0_01_10_20_3 \rangle$ and next five pulses invert the 
populations of $\vert 0_00_10_21_3 \rangle$ $\&$ $\vert 0_01_10_21_3 \rangle$. 
Figure (\ref{in2}) shows the population distribution of $\rho_{in}^2$ and the corresponding 
spectrum is shown in Fig. (\ref{ppops}b). The positive and negative transitions in Fig. (\ref{ppops}b) are in accordance with the 
population distribution of Fig. (\ref{in2}).

\begin{figure}
\begin{center}
\epsfig{file=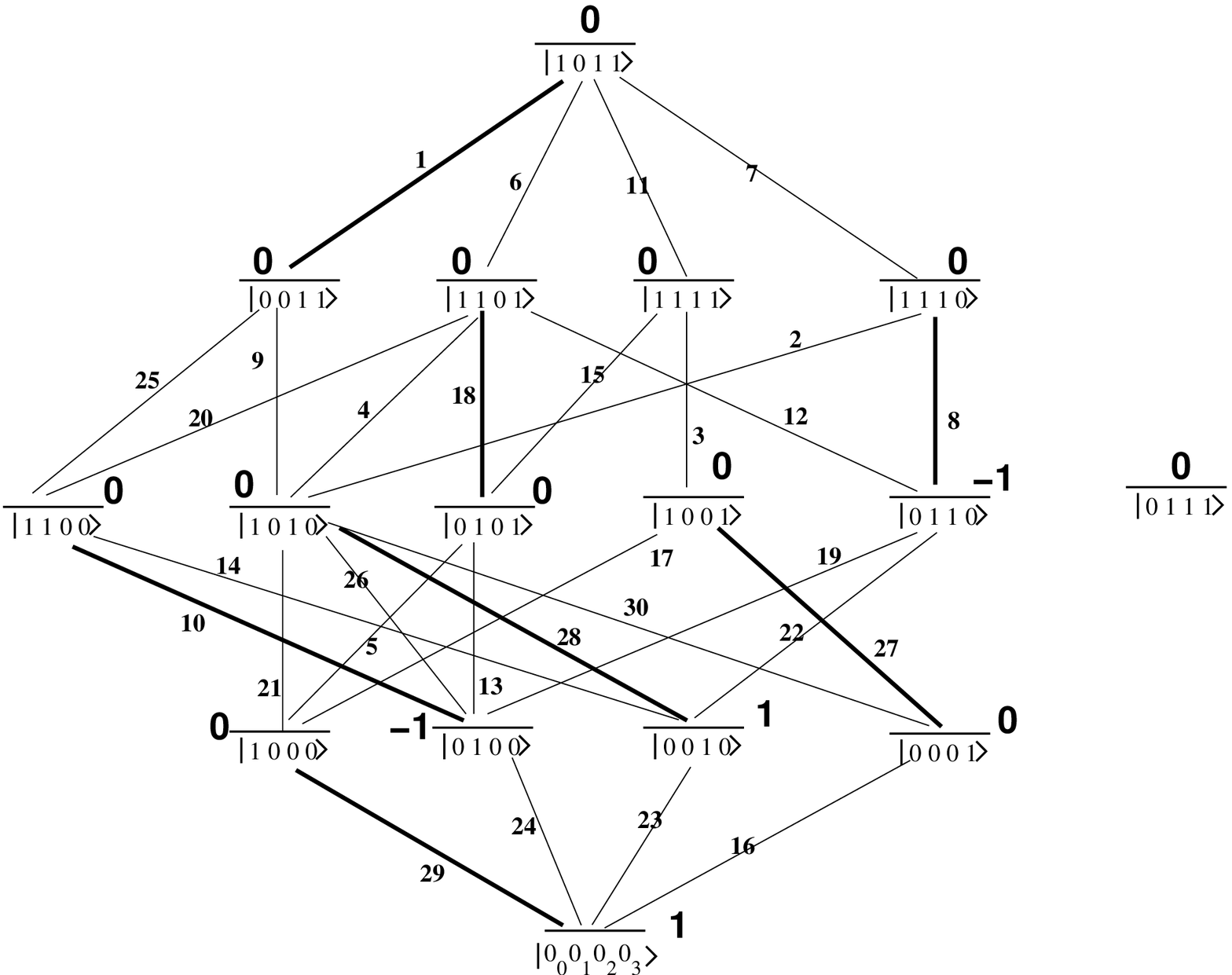,width=16cm,angle=0}
\caption{Energy level diagram of Fig. (\ref{energy}), where the populations of various states correspond to 
$\rho_{in}^3$ (Eq. \ref{5eqn21}). The
transitions represented by dark lines, correspond to
ancilla qubit}
\label{in3}
\end{center}
\end{figure}

Similarly one obtains, 

\begin{eqnarray}
\rho_{in}^3=\rho_{eq}-\rho_3&=&\vert 0_00_10_20_3 \rangle \langle 0_00_10_20_3 \vert+
\vert 0_00_11_20_3 \rangle \langle 0_00_11_20_3 \vert  \nonumber \\
&&-\vert 0_01_10_20_3 \rangle \langle 0_01_10_20_3 \vert-\vert 0_01_11_20_3 \rangle \langle 0_01_11_20_3 \vert, \label{5eqn21}
\end{eqnarray}

where $\rho_3$ prepared from $\rho_{eq}$ (Fig. \ref{energy}), 
by applying two selective $\pi$ pulses on transitions (24) and (22) which invert the populations of 
$\vert 0_00_10_20_3 \rangle$ $\&$ $\vert 0_01_10_20_3 \rangle$ and $\vert 0_00_11_20_3 \rangle$ $\&$ $\vert 0_01_11_20_3 \rangle$ 
respectively. 
Figure (\ref{ppops}c) shows the spectrum of $\rho_{in}^3$ which reflects the population distribution of Fig. (\ref{in3}).

{\bf (ii) Oracle:}
To search the state $\vert \psi_6 \rangle$ (Eq. \ref{5eqn17}), the unitary operator (U) of oracle, interchanges the states 
$\vert 0_0 \psi_6 \rangle$ and $\vert 1_0 \psi_6 \rangle$. 
Here U is achieved by applying a selective $\pi$ pulse on transition (18). 
The resultant states $\rho_{f}^1$, $\rho_{f}^2$ and $\rho_{f}^3$ are given by,

\begin{eqnarray}
\rho_{f}^1&=&U(\rho_{in}^1)=U(\rho_{eq})-U(\rho_1) =\rho_{in}^1 \nonumber \\
\rho_{f}^2&=&U(\rho_{in}^2)=U(\rho_{eq})-U(\rho_2) \nonumber \\
&=&U(\vert 0_00_10_20_3 \rangle \langle 0_00_10_20_3 \vert)+U(\vert 0_00_10_21_3 \rangle
\langle 0_00_10_21_3 \vert)
-U(\vert 0_01_10_20_3 \rangle \langle 0_01_10_20_3 \vert) \nonumber \\
&&-U(\vert 0_01_10_21_3 \rangle \langle 0_01_10_21_3 \vert) \nonumber \\
&=&\vert 0_00_10_20_3 \rangle \langle 0_00_10_20_3 \vert+\vert 0_00_10_21_3 \rangle \langle 0_00_10_21_3 \vert
-\vert 0_01_10_20_3 \rangle \langle 0_01_10_20_3 \vert \nonumber \\
&&-(\vert 1_01_10_21_3 \rangle \langle 1_01_10_21_3 \vert \nonumber \\
\rho_{f}^3&=&U(\rho_{in}^3)=U(\rho_{eq})-U(\rho_3)=\rho_{in}^3.   \label{5eqn22}
\end{eqnarray}

In Eq. (\ref{5eqn22}), U converts the state $\vert 0_01_10_21_3 \rangle$ of $\rho_{in}^2$ in to 
$\vert 1_01_10_21_3 \rangle$, whereas $\rho_{in}^1$ and $\rho_{in}^3$ are not effected by U, since neither of these states 
contain either $\vert 0_01_10_21_3 \rangle$ or $\vert 1_01_10_21_3 \rangle$.

\begin{figure}
\begin{center}
\epsfig{file=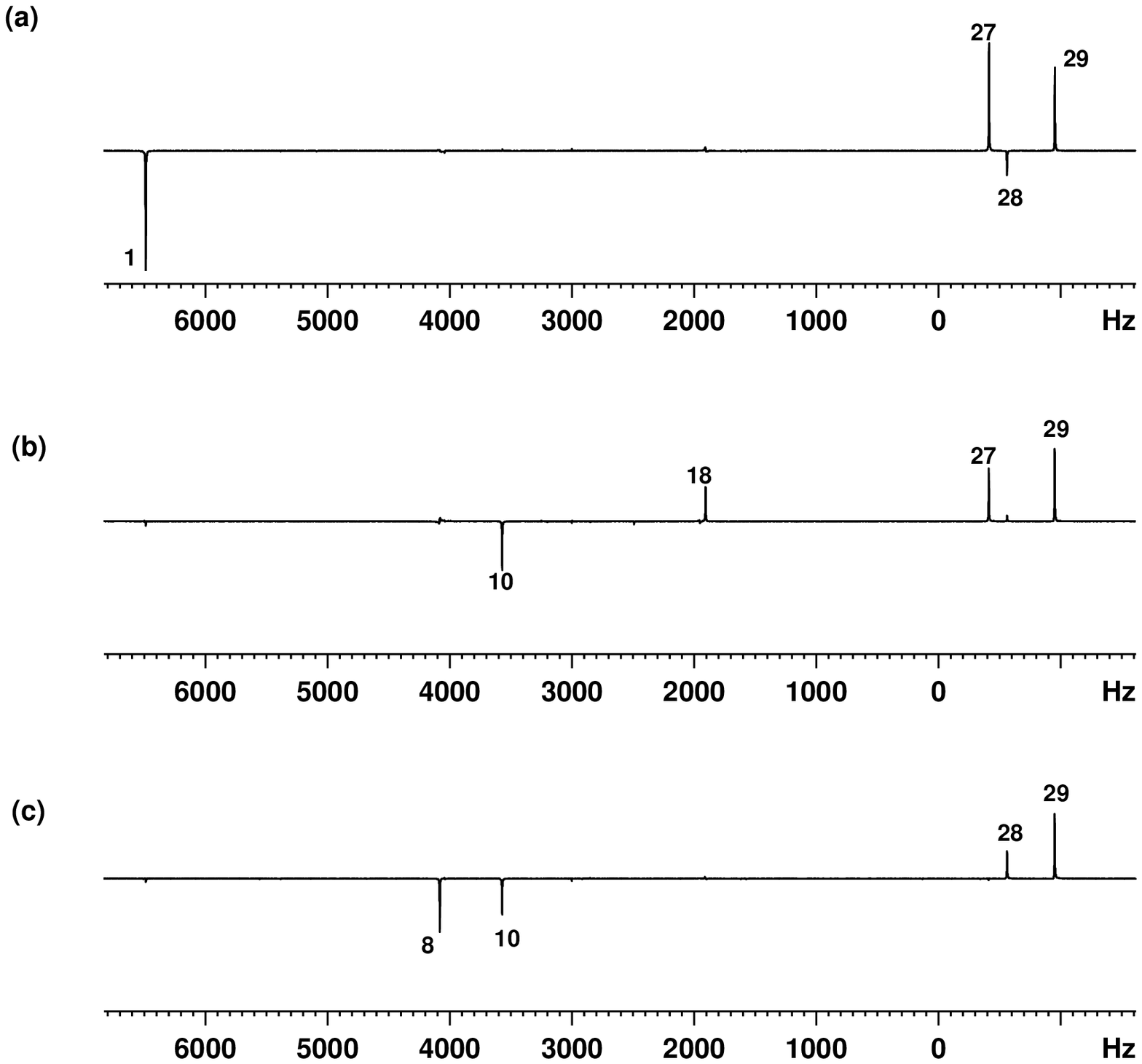,width=10cm,angle=270}
\caption{Implementation of Generalized Liouville space search algorithm on a dipolar coupled four qubit system.
The spectra of U($\rho_{eq}$), U($\rho_{1}$), U($\rho_{2}$) and U($\rho_{3}$)  (Eq. \ref{5eqn22}) are first obtained by applying a multi
frequency pulse on ancilla qubit transitions (1), (8), (10), (18), (27), (28) and (29). Then the spectra (a), (b) and (c) are 
respectively obtained by
subtracting the
spectra of U($\rho_{1}$), U($\rho_{2}$) and U($\rho_{3}$) from U($\rho_{eq}$) spectrum. (a), (b) and (c) thus respectively represent 
the ancilla qubit spectra of $\rho_f^1$, $\rho_f^2$ and $\rho_f^3$ (Eq. \ref{5eqn22}). 
Since $\rho_f^1=\rho_{in}^1$ and $\rho_f^3=\rho_{in}^3$, the signs of ancilla qubit transitions of (a) and (c) are respectively 
equivalent to that of Fig. s (\ref{ppops}a) and (\ref{ppops}c), whereas in (b) of $\rho_f^2$ the sign of transition (8) is changed 
from that of $\rho_{in}^2$ Fig. (\ref{ppops}b), due to the oracle operation. 
Thus the spectra of $\rho_f^1$ and  $\rho_f^3$ (a,c) contains equal number of positive and negative peaks, whereas the spectrum of 
$\rho_f^2$ (b) contains unequal number of positive and negative peaks. Thus the search state is $\vert 101 \rangle$.}
\label{search1}
\end{center}
\end{figure}

{\bf (iii) Measurement:}
Measurement requires a multi frequency $(\pi/2)_y$ on eight unconnected ancilla qubit transitions
of the final state ($\rho_f^k$) followed by the detection. The eight transitions of MF pulse correspond to
eight pairs energy levels,
$\vert 0_0i_1i_2i_3 \rangle$ $\Leftrightarrow$ $\vert 1_0i_1i_2i_3 \rangle$, where $i_1, i_2, i_3$ = 0 or 1.
The transition between the energy levels $\vert 0_01_11_21_3 \rangle$ and $\vert 1_01_11_21_3 \rangle$ is not observed in the 
Z-COSY assignment of Fig. (\ref{5equi}) \cite{5mahesh1}, due to 
low intensity. 
However this transition is not required, since it any way gives rise to zero intensity peak, because the final states $\rho_f^k$ 
after the oracle of search sate 
$\vert \psi_6 \rangle$ (Eq. \ref{5eqn22}) and also for remaining search states, contain 
zero populations in $\vert 0_01_11_21_3 \rangle$ and $\vert 1_01_11_21_3 \rangle$. 
Hence the multi frequency $(\pi/2)$ pulse is applied on remaining seven transitions, 
(1), (8), (10), (18), (27), (28), (29) (represented by dark lines in Fig.s \ref{in1}, \ref{in2}, \ref{in3}).
The duration of MF pulse is 70 ms, which is obtained by modulating the Gaussian pulse with seven harmonics corresponding to these seven
transitions, the phase of each harmonic is 'y'. The amplitude of each harmonic is adjusted such that the pulse gives 
maximum intensity for each transition, 
in other words, the MF pulse corresponds to a $(\pi/2)_y$ pulse on each of the seven transitions. 
The spectra of $\rho_{f}^1$, $\rho_{f}^2$ and $\rho_{f}^3$ are given in Fig.s \ref{search1} (a), (b) and (c) respectively.
In Fig.s (\ref{search1}a, c) number of positive peaks is equal to number of negative peaks, hence the first and third qubits of the 
marked state, are in state $\vert 1 \rangle$. Whereas in Fig. (\ref{search1}b), number of positive peaks is not equal to number of negative peaks,
hence second work qubit is in state $\vert 0 \rangle$. Thus the marked is $\vert 1_10_21_3 \rangle$.

It may be pointed out that, since the ancilla qubit transition 
$\vert 0111 \rangle$ $\Leftrightarrow$ $\vert 1111 \rangle$ is not observed (Fig. \ref{5equi}, \ref{energy}), one can not perform 
the Oracle operation (U), to search the state $\vert 111 \rangle$ with the present labeling scheme.

\section{5. Conclusions}
In order to increase the number of qubits, one has to exploit the dipolar couplings among the spins, in which case the spins are 
often strongly coupled to each other. Unlike weakly coupled systems, strongly coupled systems can not be directly used for implementing 
quantum algorithms. 
In this work, we generalize the Liouville space search algorithm, such that it can be implemented in weakly as well as strongly coupled 
systems. 
Experimental implementation is carried out on a strongly dipolar coupled four qubit system. 
All the steps of the algorithm are implemented by using transition selective pulses.

\pagebreak

\pagebreak

\pagebreak

\pagebreak

\pagebreak

\pagebreak

\pagebreak

\pagebreak


\begin{thebibliography}{99}

\bibitem{5ic} M.A. Nielsen , I.L. Chuang. {\it"Quantum Computation and Quantum
Information",} Cambridge University Press, Cambridge, U.K. 2000.
\bibitem{5tur} A. M. Turing, {\it On computable numbers with an application to the Entschneidungsproblem}, Proc. London Math.Soc. 
{\bf 42}, 230 (1936).
\bibitem{5church} A. Church, {\it An unsolavable problem of elementary number theory},
{\it Am. J. Math.,} Am. J. Math. {\bf 58}, 345 (1936).
\bibitem{5deujoz} D. Deutsch and R. Jozsa, {\it Rapid solution of problems by quantum computation}, 
Proc. R. Soc. Lond. A, {\bf 439}, 553 (1992).
\bibitem{5div}D. P. DiVincenzo, {\it Quantum computation}, Science {\bf 270}, 255, (1995).
\bibitem{5gr} L.K. Grover, {\it Quantum Mechanics helps in searching for a needle in haystack}, Phys. Rev. Lett. {\bf 79}, 325 (1997).
\bibitem{5shor} P. W. Shor, {\it Polynominal-time algorithms for prime factorization and discrete algorithms on 
quantum computer}, SIAM Rev, {\bf 41}, 303-332 (1999).




\bibitem{5madi} Z. L. Madi, R. Bruschweiler, and R. R. Ernst, {\it One- and two-dimensional ensemble quantum computing in
spin Liouville space}, J. Chem. Phys. {\bf 109}, 10603 (1998).
\bibitem{5bru}  R. Bruschweiler, {\it Novel Strategy for Database Searching in Spin Liouville Space by NMR Ensemble Computing}, 
Phys. Rev. Lett. {\bf 85}, 4815(2000).
\bibitem{5dhe} C. Dhelon and V. Protopopescu,
{\it Journal of Physics A: Mathematical and General} {\bf 35}, pp. L597-L604 (2002).
\bibitem{5prot} V. Protopopescu, C. DHelon, and J. Barhen {\it Journal of Physics A} {\bf 36} L399 (2003).
\bibitem{5xiao} L. Xiao, G. L. Long, H. Y. Yan, and Y. Sun, {\it Experimental realization of the Brüschweiler's algorithm in a homonuclear system}, J. Chem. Phys. {\bf 117}, 3310 (2002).
\bibitem{5long} G. L. Long,  L. Xiao, {\it Experimental realization of a fetching algorithm in a 7-qubit NMR spin Liouville space 
computer}, J. Chem. Phys. {\bf 119}, 8473 (2003).
\bibitem{5xiao1} L. Xiao, and  G. L. Long, {\it Fetching marked items from an unsorted database in NMR ensemble computing}, 
Phys. Rev. A {\bf 66}, 052320 (2002).
\bibitem{5long1} G. L. Long,  L. Xiao, {\it  Parallel quantum computing in a single ensemble quantum computer}, 
Phys. Rev. A {\bf 69}, 052303 (2004).





\bibitem{5rev1} J. A. Jones, {\it NMR quantum computation}, progress in NMR spectroscopy {\bf 38}, 325 (2001).
\bibitem{5rev2} L. M. K. Vandersypen and I. L. Chuang, {\it NMR techniques for quantum control and computation},
Review of Modern Physics {\bf 76}, 1037 (2004).
\bibitem{5ern} R. R. Ernst, G. Bodenhausen, and A. Wokaun, 
{\it Principles of Nuclear Magnetic Resonance in One and Two Dimensions, Oxford University Press (1987)}
\bibitem{5fung} B. M. Fung, {\it Use of pairs of pseudopure states for NMR quantum computing}, Phys. Rev. A  {\bf 63}, 022304 (2001).   
\bibitem{5fung1}  A. K. Khitrin and B. M. Fung, {\it Nuclear magnetic resonance quantum logic gates using quadrupolar nuclei},
J. Chem. Phys. {\bf 112}, 6963 (2000).
\bibitem{5mahesh} T. S. Mahesh, Neeraj Sinha, K. V. Ramanathan, and Anil Kumar, 
{\it Ensemble quantum-information processing by NMR: Implementation of gates and the creation of pseudopure states using dipolar coupled spins as qubits }, Phys. Rev. A {\bf 65}, 022312 (2002).
\bibitem{5mahesh1} T. S. Mahesh, Neeraj Sinha, Arindam Ghosh, Ranabir Das, N.Suryaprakash, Malcom H.Levitt, 
K. V. Ramanathan, and Anil Kumar, 
{\it Quantum information processing by NMR using strongly coupled spins}, 
Current Science {\bf 85}, 932 (2003); xxx.lanl.gov/abs/quant-ph/0212123.
\bibitem{5rana} Ranabir Das and Anil Kumar, {\it Quantum information processing by NMR using a 5-qubit system formed by dipolar coupled spin
s in an oriented molecule}, J. Magn. Reson. {\bf 170} 310 (2004).
\bibitem{5khitrin} Jae-Seung Lee and A. K. Khitrin, {\it Pseudopure state of a twelve-spin system}, J. Chem. Phys 
{\bf 122}, 041101 (2005).
\bibitem{5mahesh2} T. S. Mahesh and Dieter Suter, {\it  Quantum-information processing using strongly dipolar coupled nuclear spins}, 
Phys. Rev. A  {\bf 74}, 062312 (2006).
\bibitem{5grace} R. C. R. Grace and Anil Kumar, {\it Flip angle dependence of non-equilibrium states yielding information on 
connectivity of transitions and energy levels of oriented molecules}, J. Magn. Reson. {\bf 99}, 81 (1992).

\end{thebibliography}
\end{document}